\documentclass[12pt]{article}

\newcommand{\be}{\begin{equation}}
\newcommand{\ee}{\end{equation}}
\newcommand{\bea}{\begin{array}}
\newcommand{\ea}{\end{array}}
\newcommand{\beqa}{\begin{eqnarray}}
\newcommand{\eeqa}{\end{eqnarray}}
\newcommand{\bean}{\begin{eqnarray*}}
\newcommand{\eean}{\end{eqnarray*}}
\newcommand{\eqn}[1]{(\ref{#1})}

\newcommand{\Spin}{\mbox{\it Spin}}

\newcommand{\gapproxeq}{\lower .7ex\hbox{$\;\stackrel{\textstyle
>}{\sim}\;$}}
\newcommand{\lapproxeq}{\lower .7ex\hbox{$\;\stackrel{\textstyle
<}{\sim}\;$}}

\newcounter{appendice}

\def\thebibliography#1{{\bf REFERENCES\markboth
 {REFERENCES}{REFERENCES}}\list
 {[\arabic{enumi}]}{\settowidth\labelwidth{[#1]}\leftmargin\labelwidth
 \advance\leftmargin\labelsep
 \usecounter{enumi}}
 \def\newblock{\hskip .11em plus .33em minus -.07em}
 \sloppy
 \sfcode`\.=1000\relax}

\def\BI{{\rm 1\!l}}
\def\tauonerho{\matrix{{}\cr \tau^\rho  \cr {}^1}}
\def\tauonesig{\matrix{{}\cr \tau^\sigma  \cr {}^1}}
\def\tautworho{\matrix{{}\cr \tau^\rho  \cr {}^2}}
\def\tautwosig{\matrix{{}\cr \tau^\sigma  \cr {}^2}}
\def\tauonemu{\matrix{{}\cr \tau^\mu  \cr {}^1}}
\def\tautwonu{\matrix{{}\cr \tau^\nu  \cr {}^2}}

\def\tauonenu{\matrix{{}\cr \tau^\nu  \cr {}^1}}
\def\tautwomu{\matrix{{}\cr \tau^\mu  \cr {}^2}}

\def\xone{\matrix{{}\cr u \cr {}^1}}
\def\xtwo{\matrix{{}\cr u \cr {}^2}}
\def\xxone{\matrix{{}\cr x \cr {}^1}}
\def\xxtwo{\matrix{{}\cr x \cr {}^2}}
\def\xonep{\matrix{{}\cr u' \cr {}^1}}
\def\xtwop{\matrix{{}\cr u' \cr {}^2}}
\def\hxone{\matrix{{}\cr\hat u \cr {}^1}}
\def\hxtwo{\matrix{{}\cr\hat u \cr {}^2}}
\def\hxonep{\matrix{{}\cr\hat u' \cr {}^1}}
\def\hxtwop{\matrix{{}\cr\hat u' \cr {}^2}}

\def\gone{\matrix{{}\cr g \cr {}^1}}
\def\sgone{\matrix{{}\cr s \cr {}^1}}
\def\gonein{\matrix{{}\cr g^{-1} \cr {}^1}}
\def\sgtwo{\matrix{{}\cr s \cr {}^2}}
\def\gtwo{\matrix{{}\cr g \cr {}^2}}
\def\hgone{\matrix{{}\cr \hat g \cr {}^1}}
\def\hgtwo{\matrix{{}\cr \hat g \cr {}^2}}
\def\hhone{\matrix{{}\cr \hat h \cr {}^1}}
\def\hhtwo{\matrix{{}\cr \hat h \cr {}^2}}
\def\gbtwo{\matrix{{}\cr h \cr {}^2}}
\def\sgbtwo{\matrix{{}\cr \bar s \cr {}^2}}
\def\gbtwoin{\matrix{{}\cr h^{-1} \cr {}^2}}
\def\gbonein{\matrix{{}\cr h^{-1} \cr {}^1}}
\def\gbone{\matrix{{}\cr h \cr {}^1}}
\def\sgbone{\matrix{{}\cr \bar s \cr {}^1}}
\def\Ronetwo{\matrix{{}\cr R \cr {}^{12}}}
\def\TRonetwo{\matrix{{}\cr{\rm Tr }\cr {}^{12}}}
\def\Ronethree{\matrix{{}\cr R \cr {}^{13}}}
\def\Rtwothree{\matrix{{}\cr R \cr {}^{23}}}
\def\ronetwo{\matrix{{}\cr r \cr {}^{12}}}
\def\rtwoone{\matrix{{}\cr r \cr {}^{21}}}
\def\ronetwop{\matrix{{}\cr r' \cr {}^{12}}}
\def\rtwoonep{\matrix{{}\cr r' \cr {}^{21}}}
\def\ronethree{\matrix{{}\cr r \cr {}^{13}}}
\def\rtwothree{\matrix{{}\cr r \cr {}^{23}}}

\def\hronetwo{\matrix{{}\cr\hat r \cr {}^{12}}}
\def\hrtwoone{\matrix{{}\cr \hat r \cr {}^{21}}}
\def\hrtwothr{\matrix{{}\cr \hat r \cr {}^{23}}}
\def\hronethr{\matrix{{}\cr \hat r \cr {}^{13}}}
\def\hhonein{\matrix{{}\cr  \hat h^{-1} \cr {}^1\;}}
\def\hhtwoin{\matrix{{}\cr  \hat h^{-1} \cr {}^2\;}}
\def\Mone{\matrix{{}\cr M \cr {}^1}}

\def\Mtwo{\matrix{{}\cr M \cr {}^2}}
\def\up#1{\leavevmode \raise.16ex\hbox{#1}}

\def\sqr#1#2{{\vcenter{\vbox{\hrule height.#2pt
        \hbox{\vrule width.#2pt height#1pt \kern#1pt
          \vrule width.#2pt}
        \hrule height.#2pt}}}}

\def\BI{{\rm 1\!l}}

\begin{document}
\begin{flushright}

\baselineskip=12pt

DSF--21/2005\\
\hfill{ }\\
July 2005
\end{flushright}

\centerline{ \LARGE On the Absence of Continuous Symmetries}
\vskip .5cm
\centerline{ \LARGE  for Noncommutative 3-Spheres}

\vskip 2cm

\centerline{ {\sc  Fedele Lizzi$^{a}$, Allen Stern$^{a,b}$ and
Patrizia Vitale$^{a}$ }  }

\vskip 1cm
\begin{center}
 {\it a) Dip. Scienze Fisiche, Universit\`{a} di Napoli \emph{Federico II}\\ and INFN Sez. di
Napoli,\\ Compl. Univ. Monte S. Angelo, Napoli, 80126, Italy\\}
 {\it b) Department of Physics, University of Alabama,\\ Tuscaloosa,
Alabama 35487, USA\\{\small\tt fedele.lizzi@na.infn.it,
patrizia.vitale@na.infn.it, astern@bama.ua.edu}}
\end{center}
\vskip 2cm

\vspace*{5mm}

\normalsize
\centerline{\bf ABSTRACT}

A large class of noncommutative spherical manifolds was obtained
recently from cohomology considerations. A one-parameter family of
twisted 3-spheres was discovered by Connes and Landi, and later
generalized to a three-parameter family by Connes and
Dubois-Violette. The spheres of  Connes and Landi were shown to be
homogeneous spaces for certain compact quantum groups. Here we
investigate whether  this property can be extended to the
noncommutative three-spheres of Connes and Dubois-Violette. Upon
restricting  to quantum groups which are continuous deformations
of $\Spin(4)$ and $SO(4)$ with standard co-actions, our results
suggest that this is not the case.
 \vskip 2cm \vspace*{5mm}

\newpage

\section{Introduction}
The recent interest in noncommutative geometry~\cite{books} has
led to an on-going search for non-trivial examples of
noncommutative spaces. Noncommutative generalizations of spheres
in various dimensions are known (for a review
see~\cite{Dabrowskigarden}), but many  of these suffer from a drop
in dimensions.  Actually, the dimension of a noncommutative space
is not uniquely defined.  One choice which  uses concepts which are
natural in noncommutative geometry is the Hochschild dimension.  It
plays
an important role for the   three-parameter family of deformations
of the sphere $S^3$  introduced by Connes and Dubois-Violette  \cite{Connes:2001ef},
which generalizes the one-parameter family  discovered  previously
by Connes and Landi \cite{Connes:2000tj}.  The  Hochschild dimension of the
corresponding algebra remains constant (and equal to three) for
the deformation.  A generalization to higher dimensions is
possible  for the one-parameter subset, the so-called twisted
spheres. This particular subset has another important
characteristic. It has been shown~\cite{Varilly:2001qm,
Aschieri:2001gv} that the spheres in the one parameter subset carry a coaction of the multiparametric
quantum orthogonal groups $SO_\theta(n+1)$~\cite{Schirr}; i.e.,
they are homogeneous spaces of quantum groups. The aim of this
paper is to investigate the possibility  of  defining a group
coaction for the three-parameter 3-spheres as well.

The algebra of the  three-parameter spheres
$S^3_u$~\cite{Connes:2001ef} is
generated by hermitean operators $\hat x_\mu,\;\mu=0,1,2,3$,
subject to \be \hat x_\mu\hat x_\mu =\BI \label{untspr}\;,\ee
where  $\BI$ is the unit operator, and quadratic commutation
relations
\be
[\hat x_\mu,\hat x_\nu] =i \hat E_{\mu\nu,\,\rho\sigma} \; \hat
x_\rho \hat x_\sigma\ .\label{cdvcrs}
\ee
The constant coefficients $\hat
E_{\mu\nu,\,\rho\sigma}$ are expressed in terms of four angles
$\phi_\mu$,
\be
\hat E_{\mu\nu,\,\rho\sigma}=
\epsilon_{\mu\nu\rho\sigma}\;\frac{\sin{(\phi_\rho-\phi_\sigma)}}{
\cos{(\phi_\mu-\phi_\nu)}} \;,\ \ \ \ \mbox{\small no sum on repeated
indices}  \label{Ehat}
\ee
$\hat E_{\mu\nu,\,\rho\sigma}$  is antisymmetric in the first two
indices and symmetric in the last two indices. These commutation
relations hold provided no two angles differ by $\pi/2$. Because
they depend only on the difference of angles there are three
independent deformation parameters, and so one angle, say
$\phi_0$, can be set to zero. The Connes-Landi case has two of the
remaining angles equal with the third zero; e.g.,
$\phi_1=\phi_2=\frac\phi 2$ and $\phi_3=0$~\cite{Connes:2000tj}.

In  the search for continuous symmetries,  we shall  consider
linear, as well as  spinor, transformations.  When all angles are
set to zero we require that  the symmetries reduce to $SO(4)$ and
$\Spin(4)$ transformations, respectively. On the other hand,  we
cannot get   Lie group transformations when any of the independent
parameters are nonvanishing, since~\eqn{cdvcrs} would in general
not  be preserved. If they exist, such symmetries  should
correspond to quantum group transformations. As the Connes-
Dubois-Violette three-spheres are three-parameter deformations of
the sphere,  their  symmetries should correspond to
multiparametric  deformations of $SO(4)$ and $\Spin(4)$.
Multiparametric deformations of orthogonal groups
\cite{Schirr,mpd} can be  obtained from the standard one-parameter
quantum group by applying a twist $F$, which depends on additional
parameters $q_{ab}$, to the quantum $R$ matrix.  The twist is
required to be a specific function in the universal enveloping
algebra of the Lie group under consideration. Under these
assumptions the quantum deformation becomes at most two-parametric
in the case of $SO(4)$. Therefore these kind of q-groups  cannot
be be associated with symmetries for the full three-parameter
family of noncommutative spheres.  Alternatively, it is  possible
that there  exist symmetries associated with quantum groups which
are not deformable to Lie-groups. Here, however, our primary focus
will be on continuous deformations of $\Spin(4)$ and $SO(4)$.  We
are then justified in looking at the limit of small angles where
the search for continuous symmetries is considerably simplified.
This is the commutative limit, where
 the noncommutative sphere goes to  $S^3$, with $\hat x_\mu$ going to real commuting
coordinates $x_\mu$, $x_\mu x_\mu =1$, and the noncommutativity
gets replaced by a nontrivial Poisson structure on $S^3$. The
commutative limit of the quantum group associated with a
continuous symmetry, if it exists, is a Lie-Poisson
group~\cite{Takhtajan}, a Lie group with a Poisson bracket on the
group manifold which is compatible with the group multiplication.

The search for  Lie-Poisson symmetry in the case of the
commutative limit of the Connes and Dubois-Violette spheres is
carried in sections 2 and 3. The Poisson brackets are recovered from (\ref{cdvcrs})
 in the limit of
small angles $\phi_\mu \rightarrow\epsilon_\mu$  \be
\{x_\mu,x_\nu\} = E_{\mu\nu,\,\rho\sigma} \; x_\rho
x_\sigma\;,\label{pbxmxn} \ee where \be E_{\mu\nu,\,\rho\sigma}=
\epsilon_{\mu\nu\rho\sigma} (\epsilon_\rho-\epsilon_\sigma)\;, \ \
\ \ \mbox{\small no sum on repeated indices} \label{clpbs} \ee Our
approach   is to  express the Poisson brackets on  $S^3$ in terms
of a constant matrix, with the intention of utilizing it as a
classical R-matrix for a Lie-Poisson group. In section 3 we
consider $\Spin(4)$ transformations.    For this one expresses the
coordinates in terms of an $SU(2)$ matrix $u$, with the
$\Spin(4)=SU(2){\times} SU(2)$ transformation given by \be
u\rightarrow u'= g u h^{-1} \;,\label{sltcnx}\ee where  $g$ and
$h$ are independent elements of $SU(2)$.      The problem is  then
to find  a Poisson structure on $\Spin(4)$ which is  compatible
with the Poisson algebra on $S^3$.    This means that both group
multiplication and the  action on  $S^3$ are Poisson maps. The
former property defines the  Lie-Poisson group.  Expressing  the
Poisson brackets on the group  in terms of a classical $R-$matrix,
which by definition satisfies the classical Yang-Baxter equations,
insures that the Jacobi identity is  satisfied.  If such a
classical $R-$matrix is found, the group is a Lie-Poisson group,
and  the space under consideration  would be a homogeneous space
of the  Lie-Poisson group. In section 2 we find a consistent
classical $R-$matrix    only
 in the Connes-Landi
limit, and thus we only get  a Poisson map of  $\Spin(4)$  in this
case. In section 3 we get the same result for $SO(4)$.   We assume
the usual linear  $SO(4)$ transformations
\be x_\mu\rightarrow {x'}_\mu =
M_{\mu\nu}x_\nu\;,\label{orthtrns}\ee   $M_{\mu\nu}$ being
$SO(4)$ matrix elements.    A candidate for the Poisson brackets of $M_{\mu\nu}$  can be written
in terms of a constant $16\times 16$ matrix, but the latter only defines a
classical $R-$matrix, i.e., satisfies the classical
Yang-Baxter equations, in the Connes-Landi limit.

After ruling out symmetries associated with continuous deformations
of $\Spin(4)$ and $SO(4)$, there still is the
 possibility of symmetries    at certain finite angles $\phi_\mu$.
This case is more difficult to analyze since it involves going to the
full noncommutative theory.
In section 4 we investigate the full noncommutative theory and search
symmetries associated with  spinor-type transformations.    We
express the algebra for the twisted three-sphere in terms of a possible quantum
R-matrix. The quantum Yang-Baxter equations should be satisfied for the
 corresponding  quantum group
algebra
 to be co-associative. This cannot be true for arbitrary continuous
deformations of the commutative sphere, since in the limit of small
 angles we recover the system of section 2.  In section 4 we further find
no finite values of $\phi_\mu$, other than those in the Connes-Landi
limit, for which the candidate R-matrix satisfies the  quantum Yang-Baxter
equations.

In section 5 we Wick rotate the system of  Connes and
 Dubois-Violette, leading to ``noncommutative
 hyperboloids'' in  Minkowski space, and repeat some of
 the previous  analysis in search of   quantum deformations of
 the Lorentz group  which have twisted hyperboloids as homogeneous
 spaces.  As before the search is only successful for a one parameter
 subset of hyperboloids, namely being the Wick rotation of
 Connes-Landi spheres.

 In section 6 we give concluding remarks and discuss the prospects
 for a more exhaustive study of the full
noncommutative theory.

\section{Poisson action of
$\Spin(4)$} \setcounter{equation}{0}

As $u$ appearing in~(\ref{sltcnx}) is in the defining
representation of $SU(2)$ it can be expressed  in terms of the
coordinates according to \be u=x_\mu\tau^\mu \;,\label{clsx}\ee
where $\tau^0$ is the $2{\times} 2$ identity matrix
$\tau^0=\BI_{2{\times} 2}$ and $\tau^i$ are $i$ times the Pauli
matrices, $\tau^i=i\sigma^i,\;i=1,2,3$.  $\tau^\mu$ satisfy
\be\frac12 {\rm Tr}(\tau^\mu\tau^\nu)=\eta^{\mu\nu}={\rm
diag}(1,-1,-1,-1)\label{mnkmtrxc}\ee  $u$ has real trace and the traceless
part is anti-hermitean. Hermitean conjugation corresponds to a parity
transformation.

We next show that the  Poisson brackets~(\ref{pbxmxn}) can be
written in the form
\be
\{\xone,\xtwo\}=\xtwo r\xone -\xone r\xtwo\;,\label{pbitoxm} \ee
where by $\xone$ and $\xtwo$ is meant $u\otimes\BI$ and
$\BI\otimes u$ respectively, and $r$ is a $4{\times} 4$ matrix
which we need to determine\footnote{For a more general starting
ansatz see the end of this section.}. From
\be u u^\dagger =u^\dagger u =\BI_{2{\times} 2}\label{uuntry}\;,\ee $r$ must be
hermitean.  From the antisymmetry of the Poisson bracket $r$
should be invariant under exchange of the two tensor product
spaces; i.e., $r=\ronetwo=\rtwoone$.  Finally in order to recover
the Poisson brackets~(\ref{pbxmxn}) from~(\ref{pbitoxm}) $r$
should   satisfy
\be \tautworho r \tauonesig -\tauonerho r \tautwosig +\tautwosig r
\tauonerho -\tauonesig r \tautworho =2\;E_{\mu\nu,\,\rho\sigma}
 \tauonemu\tautwonu \label{cndtnnr}\ee
The solution (up to a term proportional to the identity matrix
$\BI_{4{\times} 4}= \tau^0
{\times}
\tau^0$)  is
\be
 r=\kappa_1\;\tau^1{\otimes}\tau^1\;+\;\kappa_2
\;\tau^2{\otimes}\tau^2\;+\;\kappa_3\;\tau^3{\otimes}\tau^3\;,
\label{rfcdv}\ee where $\kappa_i$ are given by   \beqa
 \kappa_1&=&\frac12 (-\epsilon_0-\epsilon_1+\epsilon_2+\epsilon_3)\cr& &\cr
\kappa_2&=&\frac12 (-\epsilon_0+\epsilon_1-\epsilon_2+\epsilon_3)\cr& &\cr
\kappa_3&=&\frac12 (-\epsilon_0+\epsilon_1+\epsilon_2-\epsilon_3)\label{dfkpa}
\eeqa

   The standard action of $\Spin(4)$ on $S^3$ is
(\ref{sltcnx}), where  $g$ and $h$ are independent
elements of $SU(2)$
 in the defining representation.  The Poisson algebra
~(\ref{pbitoxm}) is not preserved under this action.  Instead,
~(\ref{pbitoxm}) goes to \be \{\xonep,\xtwop\}=\xtwop \ronetwop
\xonep -\xonep \rtwoonep \xtwop\;,
 \ee
where \be \ronetwo' = \gone\gbtwo\; r\; \gbtwoin\gonein  \ee
 On the other
 hand, if we can consistently assign the following Poisson structure to $\Spin(4)$:
\be \{ \gone,\gbtwo \}
= -[r,  \gone\gbtwo ]   \qquad\qquad
 \{\gone,\gtwo\}=\{\gbone,\gbtwo\}=0\; \;,\label{pbsfrgagd}\ee the
 brackets~(\ref{pbitoxm}) are invariant in the sense that
~(\ref{sltcnx}) is a Poisson map. For these Poisson brackets to be
 consistent we need that they are
 antisymmetric and satisfy the Jacobi identity.  This means that
$r$ should satisfy  the classical Yang-Baxter equations
\be [\ronetwo,\ronethree +\rtwothree]+[\ronethree,\rtwothree]=I'\;,\ee
where $I'$ is an adjoint invariant for $\Spin(4)$.  Note that  the
classical Yang-Baxter equations did not have to be satisfied for
(\ref{pbitoxm}) to be consistent with the Jacobi identity.
 It is easily
seen that the classical Yang-Baxter and hence the Jacobi identity for (\ref{pbsfrgagd})
are satisfied  when all $\kappa_i$  but one vanish.
This corresponds to  two angles being equal while the third is
zero, i.e. the Connes-Landi case. {\it Moreover, the
Yang-Baxter equations and Jacobi identity are only satisfied in
this case, and thus only then do $g$ and $h$ generate a Lie-Poisson group.} In that case we can introduce spinors $\psi =
\pmatrix{\psi_1 \cr
  \psi_2}$ and  $\bar\psi = \pmatrix{\bar\psi_1 & \bar\psi_2}$ with
Poisson brackets  \be \{\psi_a,\bar\psi_b\}= r_{ad,cb}
\;\psi_c\bar\psi_d \label{spnrspb}\;,\ee for which
\be \psi\rightarrow \psi' = g \psi\;,\qquad \bar\psi\rightarrow
\bar\psi' = \bar \psi h^{-1} \ee  will be  a Poisson map.   Then
the Poisson algebra for $\psi\bar \psi$ is identical to that for  $u$
in~(\ref{pbitoxm}).

Concerning \eqn{pbitoxm},  we could start with the  most general
ansatz which is
linear in both $\xone$ and $\xtwo$:
\be
\{\xone,\xtwo\}=r^{(1)}\xone\xtwo +\xone\xtwo r^{(2)}- \xtwo
r^{(3)}\xone -\xone r^{(4)}\xtwo\;,\label{mgpbitoxm} \ee where we have
introduced  four $4{\times}4$ matrices $r^{(A)},\;A=1,2,3,4$.
 Now
(\ref{sltcnx}) is a Poisson map when~(\ref{pbsfrgagd}) is
generalized to
\beqa \{\gone,\gtwo\}&=&[r^{(1)},\gone\gtwo]\cr
\{\gbone,\gbtwo\}&=&[r^{(2)},\gbone\gbtwo]\cr
 \{ \gone,\gbtwo \}
&=& [r^{(3)},  \gone\gbtwo ]\cr
 \{ \gbone,\gtwo \}
&=& [r^{(4)},  \gbone\gtwo ]   \;,\label{mgpbsfrgagd}\eeqa The
matrices $r^{(A)},\;A=1,2,3,4,$ are not fully determined from the
Poisson brackets (\ref{pbxmxn}) of the coordinates.  The ambiguity
can be fixed  once one imposes the requirements that the Poisson
brackets for the matrix elements of $g$ and $h$ be antisymmetric
and consistent with ${\rm det}g={\rm det}h=1$.  However, it can be
shown, that then the brackets (\ref{mgpbitoxm})
and~(\ref{mgpbsfrgagd})  collapse to~(\ref{pbitoxm}) and
(\ref{pbsfrgagd}), and so the previous conclusions  apply.

\section{Poisson action of
$SO(4)$}\setcounter{equation}{0}

From the $\Spin(4)$ transformations~(\ref{sltcnx}) we can
construct the corresponding $SO(4)$ transformations~(\ref{orthtrns}), and since the
former defines a Poisson map in the Connes-Landi case so does the
latter.    In that case $SO(4)$ matrix elements
$M_{\mu\nu}$ are expressed as  quadratic functions of group elements $g$ and $h$
\be M_{\mu\nu}(g,h)=\frac 12 {\rm Tr}(\tau_\mu g\tau^\nu h^{-1})
\;,\label{sofmes}\ee where the indices on $\tau$ are raised and lowered with the
Minkowski metric~(\ref{mnkmtrxc}).  More generally, if we don't make
assumptions like
(\ref{sofmes}), it may be  possible to find a Poisson
map  of a group even when   no Poisson
map is induced by its covering group.  However,  we
find that not to be the case for $SO(4)$ acting on the noncommutative 3-sphere, i.e., like $\Spin(4)$, $SO(4)$  has
a Poisson action only in the Connes-Landi case.

  The Poisson
brackets~(\ref{pbxmxn})
 are not preserved under the action~(\ref{orthtrns}) of $SO(4)$.  Rather they are
 transformed to  \be
\{x'_\mu,x'_\nu\} = E'_{\mu\nu,\,\rho\sigma} \; x'_\rho
x'_\sigma\;,\ee where\be E'_{\mu\nu,\,\rho\sigma}=  M_{\mu\alpha}
M_{\nu\beta}M_{\rho\gamma}M_{\sigma\delta}\;E_{\alpha\beta,\gamma\delta}
\ee
 On the other
 hand, it may be possible to have  Poisson brackets on
 $SO(4)$ which define a Lie-Poisson group and make~(\ref{orthtrns})  a Poisson map.   The  Poisson brackets are
 required to satisfy
\be
 \biggl(\{M_{\mu\gamma},M_{\nu\delta}\}-E_{\mu\nu,\,\rho\sigma}M_{\rho\gamma}M_{\sigma\delta}
 +M_{\nu\sigma}M_{\mu\rho}E_{\rho\sigma,\gamma\delta}\biggr) x_\gamma
 x_\delta =0\;,\label{cfmtblp} \ee in addition to antisymmetry and the Jacobi identity.
(\ref{cfmtblp}) is  solved by \be \{ \Mone,\Mtwo \} = [R,  \Mone\Mtwo ]
\;,\label{pbsfrM}\ee where \be R_{\mu\nu,\,\rho\sigma} =
E_{\mu\nu,\,\rho\sigma}+ A_{\mu\nu,\,\rho\sigma}\;,\ee  and $A$ is
antisymmetric in the last two indices,
$A_{\mu\nu,\,\rho\sigma}=-A_{\mu\nu,\,\sigma\rho}$.  From the
requirement that $M$ is orthogonal it follows that $R$ should be a
symmetric matrix
$R_{\mu\nu,\,\rho\sigma}=R_{\rho\sigma,\,\mu\nu}$. This then fixes
$A_{\mu\nu,\,\rho\sigma}=E_{\rho\sigma,\,\mu\nu}$, and hence $R$
becomes
\be
R_{\mu\nu,\,\rho\sigma}= \epsilon_{\mu\nu\rho\sigma}
(\epsilon_\mu-\epsilon_\nu+\epsilon_\rho-\epsilon_\sigma) \ \ .
\label{rfrsof}
\ee
Antisymmetry of the Poisson bracket follows since $R$ is
antisymmetric under exchange of the tensor product spaces. 
$R$  can be expressed in terms of tensor products of $SO(4)$
generators $J_i$ and $K_i,\;i=1,2 ,3$, written in the defining
representation:$$ J_1=\frac1{2\sqrt{2}}\pmatrix{0& 1& 0& 1\cr-1&
0&
      1& 0\cr 0& -1& 0& 1\cr-1& 0&-1&      0}\qquad K_1=\frac1{2\sqrt{2}}\pmatrix{
0&-1&1&0\cr1&0&0&1\cr-1&0&0&1\cr0&-1&-1&0}
$$ $$ J_2=\frac1{2\sqrt{2}}\pmatrix{0& 1& 0& -1\cr-1& 0&
      -1& 0\cr 0& 1& 0& 1\cr1& 0&-1&      0}\qquad K_2=\frac1{2\sqrt{2}}\pmatrix{
0&1&1&0\cr-1&0&0&1\cr-1&0&0&-1\cr0&-1&1&0}
$$ \be J_3=\frac1{2}\pmatrix{0& 0& -1& 0\cr 0& 0&
      0&1\cr 1& 0& 0& 0\cr0& -1&0&      0}\qquad K_3=\frac1{2}\pmatrix{
0&0&0&-1\cr0&0&1&0\cr 0&-1&0&0\cr1&0&0&0}
\ee
They satisfy
\beqa [J_i,J_j]&=&\epsilon_{ijk} J_k\cr\cr  [K_i,K_j]&=&\epsilon_{ijk} K_k\cr\cr
[J_i,K_j]&=&0 \eeqa Defining
$ J_\pm = \sqrt{2} (J_1\pm J_2)$ and $ K_\pm = \sqrt{2} (K_1\pm K_2)$, $R$ can be written
\be
 R\;\;=\;\;2\kappa_1\;
(J_- {\otimes}K_3-K_3\otimes J_-)\;\;+\;\;2\kappa_2\; (K_+
{\otimes}J_3-J_3\otimes K_+)\;\;+\;\;\kappa_3\; (K_-
{\otimes}J_+-J_+\otimes K_-)\;\;\;, \label{capR}\ee where $
\kappa_i$ are again given by~(\ref{dfkpa}). Finally we need to
check that~(\ref{pbsfrM}) satisfies the Jacobi identity, or
equivalently that $R$ satisfies the  classical Yang-Baxter
equations:
\be [\Ronetwo,\Ronethree +\Rtwothree]+[\Ronethree, \Rtwothree]=I\;,\label{cjfsrc}\ee
where $I$ is an adjoint invariant for $SO(4)$.  From
$[J_i,K_j]=0$,  it is easily seen that~(\ref{cjfsrc}) is satisfied
when all $\kappa_i$  but one vanish.   Thus the Jacobi
identity is satisfied when two angles are equal and the third is
zero.  {\it Just as with $\Spin(4)$, the Yang-Baxter equations and
Jacobi identity are only satisfied in this case, and thus
$[M_{\mu\nu}]$ generate a Lie-Poisson group only in this case.} It
can be checked that the results in this case agree with the lowest
order commutation relations in~\cite{Aschieri:2001gv}.

Above we have argued that we can consistently define a Poisson
algebra for the $SO(4)$ matrices  only in the Connes-Landi case.
This algebra is obtainable from the  Poisson algebra (\ref{pbsfrgagd})
  on  $\Spin(4)$ using (\ref{sofmes}):
\beqa  \{ M_{\mu\nu},M_{\rho\sigma}\}(g,h) &=& \frac 14 \TRonetwo
{\matrix{{}\cr \tau_\mu  \cr {}^1}}{\matrix{{}\cr \tau_\rho  \cr
 {}^2}}\{\gone\tauonenu\gbonein, \gtwo\tautwosig\gbtwoin\}\cr &=& \frac 14 \TRonetwo
{\matrix{{}\cr \tau_\mu  \cr {}^1}}{\matrix{{}\cr \tau_\rho  \cr
 {}^2}}
\biggl(\gone\tauonenu (\gtwo r \gbonein - \gbonein r \gtwo )\tautwosig\gbtwoin
\cr& &\quad  +\; \gtwo\tautwosig (\gbtwoin r \gone -\gone r
 \gbtwoin)\tauonenu\gbonein\biggr)
\;\;\;,\eeqa
where $\TRonetwo$ means a trace over both tensor product spaces.
 We then recover the expression
(\ref{pbsfrM}) with $R$ given by
\be R_{\mu\nu,\,\rho\sigma}= \frac14 \TRonetwo  [\tau_\rho \otimes
\tau^\nu, \tau^\mu\otimes\tau_\sigma]\; r\;\;\ee  After
substituting in the general expression  for $r$ given
in~(\ref{rfcdv}) one then gets~(\ref{rfrsof}).

\section{Quantum \Spin(4) Transformation}
\setcounter{equation}{0}

Here we generalize  to the full noncommutative theory with the
goal of searching for symmetries of the noncommutative
three-sphere occurring at finite angles $\phi_\mu$.   For
simplicity we restrict  to spinor-type transformations thereby
generalizing the discussion of Section~2.

We begin by replacing the  $2{\times} 2$   matrix $u$  by another $2{\times} 2$   matrix $\hat
u $, the latter having noncommuting matrix  elements.
 The property of unitarity
\be \hat u \hat u^\dagger =\hat u^\dagger \hat u =\BI_{2{\times}
  2}\;,\label{quuntry}\ee  can be maintained  although $\hat u$ does
not have to have unit determinant.  For this~(\ref{clsx}) should
be generalized to
\be
\hat u=\hat x_\mu \;e^{i\phi_\mu}\tau^\mu \ \ . \label{qntmx}
\ee
The sum over $\mu=0,1,2,3$ is assumed.  The unitarity condition
was shown~\cite{Connes:2001ef} to be consistent with the
commutation relations~(\ref{cdvcrs}).  We next show that the
commutation relations can be expressed as
\be
\hxone \hat r \hxtwo = \hxtwo \hat r \hxone\label{critoxm}
\ee for some $4{\times} 4$ matrix $\hat r$.
These relations are invariant under interchange of the two tensor
product spaces provided $\hat r$ is invariant  under interchange
of the two tensor product spaces; i.e., $\hat r
=\hronetwo=\hrtwoone$.   In the limit of small angles
$\phi_\mu\rightarrow \epsilon_\mu$, $ \hat r$ should reduce to $
\BI_{4{\times} 4} + ir\;,$ with $r$ given in (\ref{rfcdv}),  for then the commutator of $\hxone$
with $\hxtwo$ goes to $i$ times the Poisson bracket
in~(\ref{pbitoxm}):
\be \hxone\hxtwo -\hxtwo\hxone \rightarrow i\hxtwo r\hxone -i\hxone r\hxtwo
\ee
So the task is to find $\hat r$ so that the commutation relations
(\ref{critoxm}) agree with~(\ref{cdvcrs}).  Substituting
(\ref{qntmx}) into~(\ref{critoxm}), we get
\be
e^{i(\phi_\rho+\phi_\sigma)}(\tautworho \hat r \tauonesig -
\tauonerho \hat r \tautwosig) \hat x_\rho\hat x_\sigma =0\;,
\label{tauxtau} \ee where the sum over indices is again assumed.
We cannot equate coefficients of $ \hat x_\rho\hat x_\sigma$ to zero
because they are not all independent. Rather from~(\ref{cdvcrs})
they are related by \be \hat x_\rho\hat x_\sigma=\frac12\biggl(
S_{\rho\sigma,\,\alpha\beta}+ i \hat E_{\rho\sigma,\alpha\beta}
\biggr) \hat x_\alpha\hat x_\beta \label{dcmpsxsq}\;, \ee where
\be
S_{\rho\sigma,\,\alpha\beta}=\delta_{\alpha\rho}\delta_{\beta\sigma}
+\delta_{\alpha\sigma}\delta_{\beta\rho}\;. \label{cpsdf} \ee and
$\hat E_{\rho\sigma,\,\alpha\beta}$ is given in~(\ref{Ehat}). Both
$ S_{\rho\sigma,\,\alpha\beta}$ and $\hat
E_{\rho\sigma,\,\alpha\beta}$ are symmetric in the last two
indices.~(\ref{dcmpsxsq}) then says that all quadratic
combinations of $\hat x_\mu$ can be expressed in terms of just the
symmetric ones. If we substitute into~(\ref{tauxtau}) we can then
equate coefficients of all the  symmetric combinations of $ \hat
x_\alpha\hat x_\beta$ to zero.  The result is a
generalization of~(\ref{cndtnnr}):
$$ \tautworho \hat r \tauonesig -\tauonerho \hat r \tautwosig
+\tautwosig \hat r
\tauonerho -\tauonesig \hat r \tautworho $$\be  = -ie^{i(\phi_\mu+\phi_\nu
  -\phi_\rho -\phi_\sigma)}\;\hat E_{\mu\nu,\,\rho\sigma}(  \tautwomu
\hat r \tauonenu -\tauonemu \hat r \tautwonu)\ee  Up to an overall
constant factor, it is solved by \be
\hat r =\frac12 \pmatrix{e^{2i(\phi_0-\phi_2)} +  e^{2i(\phi_0-\phi_1)}&0&0&
e^{2i(\phi_0-\phi_2)} -  e^{2i(\phi_0-\phi_1)} \cr 0&
  e^{2i(\phi_0-\phi_3)}+1& e^{2i(\phi_0-\phi_3)}-1&0 \cr 0&
  e^{2i(\phi_0-\phi_3)}-1& e^{2i(\phi_0-\phi_3)}+1&0 \cr
  e^{2i(\phi_0-\phi_2)} -  e^{2i(\phi_0-\phi_1)} &0&0 &
  e^{2i(\phi_0-\phi_2)} +  e^{2i(\phi_0-\phi_1)}\cr}\;\;, \ee or equivalently,
\be
 \hat r=\BI_{4{\times} 4} +
 i\hat\kappa_\mu\;\tau^\mu{\otimes}\tau^\mu\;,\label{qrmtrx}\ee with a
 sum over $\mu$ and\beqa \hat \kappa_0 &=&\;\;\frac i4 (3 -
 e^{2i(\phi_0-\phi_1)} - e^{2i(\phi_0-\phi_2)} -
 e^{2i(\phi_0-\phi_3)}) \cr& &\cr
\hat \kappa_1 &=&-\frac i4 (1+ e^{2i(\phi_0-\phi_1)} -
e^{2i(\phi_0-\phi_2)} - e^{2i(\phi_0-\phi_3)})\cr& &\cr\hat
\kappa_2 &=&-\frac i4 (1 - e^{2i(\phi_0-\phi_1)} +
e^{2i(\phi_0-\phi_2)} - e^{2i(\phi_0-\phi_3)})\cr & &\cr\hat
\kappa_3 &=&-\frac i4 (1 - e^{2i(\phi_0-\phi_1)} -
e^{2i(\phi_0-\phi_2)} + e^{2i(\phi_0-\phi_3)})\eeqa The
expressions for $ \hat \kappa_i,\;i=1,2,3$,  reduce $\kappa_i$
to~(\ref{dfkpa}) in the limit of small angles $\phi_\mu\rightarrow
\epsilon_\mu$, while  $ \hat \kappa_0$ is arbitrary in the limit.

 The relations~(\ref{critoxm}) are not invariant under   $\Spin(4)$.
 Alternatively, we can  try to define  a deformation of $\Spin(4)$, parametrized by two
 nonsingular  $2{\times} 2$  matrices $\hat g$ and $\hat h$ with
 noncommuting matrix elements with an involution.   The
 co-action on
 $\hat u$  is
\be \hat u\rightarrow \hat u'= \hat g \hat u \hat h^{-1}
 \;\label{qsltcnx}\ee    In order to preserve~(\ref{quuntry}) we demand
 that $\hat g$ and $ \hat h $ are unitary. Transformation~(\ref{qsltcnx}) preserves
 the commutation relations
(\ref{critoxm}) provided
\be \hat r \hgone\hhtwo =\hhtwo\hgone \hat r \qquad\qquad
 [\hgone,\hgtwo]=[\hhone,\hhtwo]=0\;\label{qcrbtnghhh} \ee
This is easily shown.  Under the co-action, the left-hand-side of
~(\ref{critoxm}) transforms to \beqa
 \hxonep \hat r \hxtwop &=& \hgone
 \hxone \hhonein \hat r \hgtwo \hxtwo \hhtwoin \cr
 &=& \hgone
 \hxone  \hgtwo \hat r\hhonein  \hxtwo \hhtwoin
\cr
 &=&\hgtwo \hgone
 \hxone   \hat r \hxtwo \hhtwoin \hhonein
\cr
 &=&\hgtwo \hgone  \hxtwo
   \hat r \hxone  \hhtwoin \hhonein \cr
 &=&\hgtwo   \hxtwo\hgone
   \hat r \hhtwoin \hxone  \hhonein \cr
 &=&\hgtwo   \hxtwo
\hhtwoin   \hat r\hgone  \hxone  \hhonein\;\; =\;\; \hxtwop \hat
 r\hxonep
\label{nvprg}
\eeqa

In order for the algebra generated by $\hat g$ and $\hat h$ to be associative it is
necessary for $\hat r$ to satisfy the
quantum Yang-Baxter equations:
\be \hronetwo\hronethr\hrtwothr =\hrtwothr\hronethr\hronetwo\label{qybe}\ee
On the other hand, the
quantum Yang-Baxter equations did not have to be satisfied for   the algebra generated by $\hat u$  to be associative.
Finally substitute~(\ref{qrmtrx}) into (\ref{qybe}). As in the infinitesimal cases,
we get an identity only  when two angles are equal and the
third is zero. Hence the spinor-type
transformations~(\ref{qsltcnx})  correspond to symmetries only in
the Connes-Landi case. For example,  choose $\phi_0=\phi_4 = 0$
and $\phi_1=\phi_2 =\phi/2$.  Then $\hat r$ simplifies to ${\rm
diag}(q,1,1,q)$, where $q= e^{-i\phi}$, and the
condition~(\ref{untspr}) is equivalent to
\be {\rm det}_q \hat u\equiv \hat u_{11}\hat u_{22}-q \hat u_{12}\hat u_{21}=1
\;\ee  This condition is preserved under~(\ref{qsltcnx}) provided
det$\hat g\;$det$\hat h^{-1} =1$, after using~(\ref{critoxm})
along with the commutation relations for matrix elements of $\hat
g$ with $\hat h^{-1}$.  It can be checked that both det$\hat g$
and det$\hat h^{-1}$ are Casimirs of the algebra
  and hence can be set to one.
From~(\ref{qsltcnx})
 one can obtain the left co-action of the coordinates \be \hat x_\mu\rightarrow \hat{x'}_\mu =
{\hat M_{\mu\nu}}\hat x_\nu\;,\label{qorthtrns}\ee with
\be \hat M_{\mu\nu}(\hat g,\hat h)=\frac 12
e^{i(\phi_\nu -\phi_\mu)}\; {\rm Tr}(\tau_\mu \hat
 g\tau^\nu \hat  h^{-1})  \ee The commutation relations for $\hat
 M_{\mu\nu}$ are then determined from the commutation relations for matrix elements of $\hat g$
with $\hat h^{-1}$.

\section{Noncommutative hyperboloids}
\setcounter{equation}{0}

The sphere of  Connes and Dubois-Violette can be Wick rotated to
Minkowski space.  The result is a three-parameter family of
``noncommutative hyperboloids''. We can then repeat the previous
analysis and search for quantum deformations of the Lorentz group
which have noncommutative hyperboloids as homogeneous spaces.  For
simplicity, we only  examine the first order system and write it
in spinor notation.  The result is that  there is a Lie-Poisson
action of a  Lie-Poisson group acting on a one parameter subgroup
of noncommutative hyperboloids, namely the Wick rotated version
Connes-Landi spheres.

The Wick rotation of the Poisson structure~(\ref{pbxmxn}) is \be
\{x_\mu,x_\nu\} = E_{\mu\nu,\,\rho\sigma} \; x^\rho
x^\sigma\;,\label{wrpbxmxn}\ee where $ E_{\mu\nu,\,\rho\sigma}$
are again given in terms of three independent parameters by
(\ref{clpbs}) and the  indices of $x$ are raised and lowered by
the Minkowski metric $\eta_{\mu\nu}={\rm diag}(-1,1,1,1)$.
$\eta_{\mu\nu}x^\mu x^\nu$ is a Casimir for the Poisson algebra
and so we can restrict to a  hyperboloid. In the quantized theory,
the time component $x_0$ will be noncommuting for any nontrivial
values of the parameters.

 The algebra can be re-expressed in
terms of a hermitean matrix $x=x_\mu\sigma^\mu$, where $\sigma^0$
is the $2{\times} 2$ identity matrix and $\sigma^i,\;i=1,2,3,$ are
the Pauli matrices.  The Poisson brackets~(\ref{wrpbxmxn}) can be
written as

\be
\{\xxone,\xxtwo\}=i\xxtwo  r\xxone -i\xxone
r\xxtwo\;,\label{wrpbitoxm} \ee  using the definition of $r$
in~(\ref{rfcdv}).
 The Casimir is now expressed
as ${\rm det}\; x$, the latter being invariant under $SL(2,C)$
transformations
\be x \rightarrow x' = s x s^\dagger\;,\qquad s\in SL(2,C)\label{wrsltcnx}\ee
 The Poisson algebra
  is not preserved under this action, but if we can assign the following Poisson structure to $SL(2,C)$:
\be \{ \sgone,\sgbtwo \}
=-i [ r,  \sgone\sgbtwo ]   \qquad\qquad
 \{\sgone,\sgtwo\}=\{\sgbone,\sgbtwo\}=0\; \;,\ee where $\bar
 s={s^\dagger}^{-1}$, then the
 brackets~(\ref{wrpbitoxm}) are invariant in the sense that
~(\ref{wrsltcnx}) is a Poisson map. Once again for  consistency we
 need to check the Jacobi identity, or equivalently the classical
 Yang-Baxter equations for $ r$.  But as before these conditions
 are  only satisfied when all but one  $\kappa_i$ vanish, and so we only
 get a consistent deformation of $SL(2,C)$
 in the Connes-Landi limit. For  a
 classification of consistent quantum deformations of $SL(2,C)$ see~\cite{deAzcarraga:1995hh}.

\section{Concluding Remarks}
\setcounter{equation}{0}

Our search for quantum deformations of $\Spin(4)$ and  $SO(4)$ for
which the noncommutative three-spheres of Connes and
Dubois-Violette are  homogeneous spaces, and which have a smooth
commutative limit in the Lie-Poisson sense, has yielded only the
known symmetries of twisted Connes-Landi spheres. The question
arises as to whether a more involved analysis can yield any other
quantum deformations, possibly without a smooth commutative limit.
Other possibilities which  are currently under investigation would
allow for more general Poisson structures on the group at the
Poisson level, or commutation relations at the quantum level. In
this regard, although the brackets~(\ref{pbsfrgagd}), if they
could have  been consistently defined, would have assured
that~(\ref{sltcnx}) is a Poisson map, other possibilities for the
Poisson structure on $g$ and $h$ can be explored.  For example,
one might try dropping the relations $\{\gone,\gtwo\}=
\{\gbone,\gbtwo\}=0$.  The difficulty is to make this consistent
with the requirement that (\ref{sltcnx}) be a Poisson map.
Moreover, at the quantum level, one can consider generalizing the
commutation relations~(\ref{qcrbtnghhh}), possibly dropping  $
[\hgone,\hgtwo]=[\hhone,\hhtwo]=0$.  The task then would be to
find the analogue of~(\ref{nvprg}).  Finally, throughout this article we have
insisted upon writing Poisson brackets and commutations relations in
terms of an R-matrix. While this is an important case, other
possibilities should be investigated for an exhaustive study.

\subsubsection*{Acknowledgements}
We are very grateful to P.~Aschieri, J.M.~Gracia-Bond\'{\i}a and
G.~Landi for many useful discussions and correspondence. This work
has been supported in part by the \emph{Progetto Nazionale di
Interesse Nazionale {\bf Sintesi}}.

\end{document}